# Visualizing GO annotations

Fran Supek[1,2,3] and Nives Skunca[4,5,6]

[1] Division of electronics, Ruder Boskovic Institute, 10000 Zagreb, Croatia
[2] EMBL-CRG Systems Biology Unit, Centre for Genomic Regulation (CRG), 08003 Barcelona, Spain
[3] Universitat Pompeu Fabra (UPF), 08003 Barcelona, Spain
[4] Department of Computer Science, ETH Zurich, Zurich, Switzerland
[5] Swiss Institute of Bioinformatics, Zurich, Switzerland
[6] University College London, London, United Kingdom
Correspondence: fran.supek@irb.hr

## Abstract

Contemporary techniques in biology produce readouts for large numbers of genes simultaneously, the typical example being differential gene expression measurements. Moreover, those genes are often richly annotated using GO terms that describe gene function and that can be used to summarize the results of the genome-scale experiments. However, making sense of such GO enrichment analyses may be challenging. For instance, overrepresented GO functions in a set of differentially expressed genes are typically output as a flat list, a format not adequate to capture the complexities of the hierarchical structure of the GO annotation labels.

In this chapter, we survey the various methods to visualize large, difficult-to-interpret lists of GO terms. We catalogue their availability — web-based or standalone, the main principles they employ in summarizing large lists of GO terms, and the visualization styles they support. These brief commentaries on each software are intended as a helpful inventory, rather than comprehensive descriptions of the underlying algorithms. Instead, we show examples of their use and suggest that the choice of an appropriate visualization tool may be crucial to the utility of GO in biological discovery.

Keywords: Gene Ontology, visualization, interpretation, redundancy, enrichment, tools

## 1 Introduction

We have entered the era of massive data sets in biology. A variety of experimental and computational techniques can produce readouts for many genes—or whole genomes—simultaneously. Moreover, we can also assign rich functional annotations to most of the genes of interest. Such a wealth of data is accompanied with challenges in interpretation.

In this chapter, we focus on methods that visualize long lists of Gene Ontology (GO) terms *(1)*. The methods we survey take as input a flat list of GO terms, often accompanied by some user-supplied measure of statistical significance or importance. Visualization methods summarize such lists to distill the most unique and/or relevant information. Finally, these methods produce various styles of visualization that can aid in the interpretation.

First, we examine the challenges related to understanding large lists of GO terms; second, we provide a systematic overview of the published methods that address these challenges; third, we discuss different visualization styles these methods use; and fourth, we give examples of use for a selection of these tools.

# 2 Understanding large lists of genes and their Gene Ontology labels

A classical example of a large biological dataset are gene expression measurements by RNA-Seq, which monitor the genome-wide changes in transcriptional regulation between experimental conditions. Typically, tens or hundreds of genes will be up- or down-regulated in response to a particular treatment. This indicates that a systems-level change in the experimental model has occurred, and it may be described by examining the common properties of the genes whose expression was altered. Do these genes participate in the same metabolic or signaling pathways? Do they perform similar biochemical functions? Do their protein products co-localize in the cell? Formally, such sets of genes are subjected to statistical tests for enrichment for various functional categories *(2)*. The gene functions tested are typically described by Gene Ontology (GO) terms *(3)*, although alternatives such as KEGG Pathways or CORUM protein complexes could be used.

Of note, such GO enrichment analyses are by no means restricted to experiments measuring changes in gene expression, nor to experimental data in general. Any list of genes for which interpretation is sought can be described using enriched GO terms and it could, for instance, derive from comparative genomics. In particular, one could perform an evolutionary analysis to look at biological roles of gene families that have expanded in a certain eukaryotic lineage, e.g., *(4)*. Similarly, a researcher may wish to describe the overall functional repertoire in a newly sequenced genome, while comparing to existing genomes of related organisms.

## 2.1 Challenges in interpreting lists of enriched GO terms

As chapter "Primer on the GO" describes, the GO is a hierarchical structure, wherein the individual terms can have not only multiple descendants, but also multiple parents; more formally, GO is a directed graph; the basic version of the GO is also a directed *acyclic* graph[1] (Figure 1). This complex structure, along with its large size—the GO has thousands of nodes—make it rather challenging to display the part(s) of the GO of interest. Such parts could be, for instance, derived from a list of GO terms found to be enriched in a gene expression experiment.

---

[1] http://geneontology.org/page/download-ontology

A further complication is that such lists of interesting GO terms tend to be large, meaning that many different biological processes or molecular functions may appear to be affected in the experiment. One reason for this is that the GO itself is designed and developed to describe nuances in gene function as exhaustively as possible; consequently, many of the GO terms will be partially redundant. For instance, many of the genes participating in "translation" (GO:0006412) are also structurally a part of "ribosome" (GO:0005840).

In addition to the inherent redundancy of the GO, responses of biological systems to experimental perturbation often genuinely involve coordinated activity of many related and/or overlapping subsystems. For example, replicating cells facing DNA damage may upregulate "nucleotide-excision repair" (GO:0006289) to help fix the errors, but at the same time resorting to "error-prone translesion synthesis" (GO:0042276) to ensure DNA replication finishes.

## 2.2 Visualizing the GO to facilitate insight and avoid biases

GO term enrichment analyses often result in lists of significant GO terms that are both long and redundant, hampering interpretation. Various methods to visualize such lists may help investigators spot dominant trends in the data, hopefully leading to novel biological insight. Such visualizations mostly operate by different ways of grouping and displaying similar GO terms together, wherein the structure of the GO defines what is similar and what is not (see *semantic similarity analysis* below). In its simplest form, this involves displaying a part of the GO hierarchy with the GO terms of interest highlighted and their parent-child relationships shown. Displaying also the user-supplied experimental data may help prioritize which GO terms, among many similar ones, are of higher interest.

We suggest that having an unbiased way to algorithmically organize GO terms derived from experimental data helps prevent unintentional biases in interpretation. If unaware of the overall semantic structure in the set of significant GO terms, the investigator may pick one or two GO terms in the list that "make sense," in terms of fitting with their expectations. By visualizing the interrelationships between the GO terms alongside the statistical support for each in the experimental data, one would help avoid focusing on outlying—and perhaps spurious—results. Similarly, one could be made aware of the common pitfall where one GO term is chosen, while other somewhat similar and equally statistically supported terms are ignored. Finally, and very importantly, a good visualization is also an effective means of presenting summaries of scientific results, whether in papers, presentations or posters.

# 3 Overview of the GO visualization-related tools

Here we will systematize and describe the currently available tools for visualizing sets of GO annotations. Additionally, we will highlight three of these tools in more detail. The tools and the underlying methods they implement can be classified thusly:

1. Interactive GO browsers. Tools for interactively browsing the entire GO and also the genes known to be annotated with chosen GO terms. Importantly, these do not take into account a user-supplied set of annotations of interest, e.g. derived from an enrichment analysis of experimental data. Visualization is typically not emphasized and not configurable. See AmiGO *(5)* and QuickGO *(6)*. Of note, OLSVis *(7)* can display other biomedical ontologies in addition to the GO.

2. Network visualization tools. These are not particular to the GO, but can display any kind of graph, including the GO or a part thereof. The visualization options are highly configurable; however, since these tools weren't designed specifically for GO, they tend to be more complicated to use. See Cytoscape *(8)*, Gephi *(9)*, and Pajek *(10)*.
   a. Of note, there are Cytoscape plugins specialized for handling groups of GO terms: EnrichmentMap *(11)* and BINGO *(12)*.

3. GO visual overlays. Tools that can visualize an interesting subset of the GO, and display some additional data about each shown GO term. Typically, this involves coloring the GO terms by the enrichments or p-values determined from user-supplied gene lists (these tools tend to also perform the GO enrichment analysis). They display the terms arranged by parent-child relationships, in a tree-like visual layout. Examples include GOrilla *(13)*, GRYFUN *(14)*, GOFFA *(15)*, and SimCT *(16)*.
   a. In addition to the GO, similar tools are available which can highlight the individual members in displayed KEGG pathways *(17)*; the pathways can also be shown in a KEGG BRITE functional hierarchy with FuncTree *(18)*.

4. Semantic similarity analysis. Tools that examine the semantic similarity (redundancy) between various GO terms, including those that are not linked by direct parent-child relationships. The similarities are used to organize a set of interesting GO terms into clusters and/or graphs, while simultaneously allowing highly redundant terms to be filtered out. The user can supply enrichments or p-values to prioritize results. Implemented in REVIGO *(19)* and RedundancyMiner *(20)*.
   a. Some provisions for this are made in g:Profiler *(21)*, which collapses similar GO terms.
   b. The Ontologizer *(22)* can perform a statistical test for enrichment that accounts for the parent-child redundancy *(23)* prior to visualizing results.

5. Emerging methods. These may involve display of the trends underlying a group of GO terms in a so-called 'tag cloud' (with text in various colors and sizes), or in a tree map (a hierarchical organization of coloured tiles), as in REVIGO *(19)* or GOSummaries *(24)*. Additionally, several tools now support the display of multiple GO enrichment analyses side-by-side; see BACA *(25)* or GOSummaries. SimCT *(16)* can display subtrees of other biomedical ontologies in addition to the GO.

## 3.1 Case studies with selected tools

**GOrilla** *(13)* is a web-based tool that can take two types of input: either a ranked list of genes or two lists, one with the target genes and the other with the background genes. As output, GOrilla produces a visualization that indicates which terms are significantly enriched.

We focus here on the enrichment analysis that takes a ranked list of genes. Briefly, the null hypothesis is that the occurrences of a GO term at various points in the ranked list are equiprobable. The lower the displayed p-value, the more probable it is for the GO term to be enriched towards the top of the list.

As an example analysis, we downloaded a dataset of transcription profiling by microarray of human peripheral blood mononuclear cells after a treatment with *Staphylococcus aureus* and incubation for different lengths of time *(26)*, obtainedfrom the Expression Atlas *(27)* at http://www.ebi.ac.uk/gxa/experiments/E-GEOD-16837. In the GOrilla web interface (http://cbl-gorilla.cs.technion.ac.il/), we set the p-value threshold to $10^{-3}$, and the remaining settings were the defaults in the tool.

The display is shown in Figure 2. Based on the colour of the boxes, the user can clearly visualize which GO terms are enriched, and the connecting lines describe their relationship to other terms in the GO graph.

**REVIGO** *(19)* analyzes—the often large—list of significant GO terms and removes the redundant terms, to further narrow the search to a set of non-redundant and highly significant GO terms. Briefly, REVIGO creates clusters of GO terms that are semantically similar, and selects one representative for each cluster.

One possible input for REVIGO is a list of GO terms with the associated p-values, such as the output list from GOrilla. Alternatively, REVIGO can take as input any other list of GO terms, with or without associated numerical values, and provide various styles of visualization: scatterplot that distributes the GO terms, represented as bubbles, in a 2D space that will put two GO terms closer together if they are more semantically similar; interactive graph that connects the user-supplied set of GO terms based on the structure of the GO hierarchy; a TreeMap where terms are clustered by colour; and a word cloud highlights the most frequent words in the names and descriptions of the GO terms.

To perform the analysis, we used the setting in GOrilla to automatically forward its GO term enrichment results as a query to the REVIGO tool. In REVIGO, we used the default settings. The results are shown in Figures 3 and 4. The various styles of the visualization highlight the GO terms that are enriched in the input dataset.

**RedundancyMiner** *(20)* is another tool to focus on non-redundant terms in a large list of enriched GO terms, producing a Clustered Image Map (CIM) as a result. It is part of a larger pipeline: RedundancyMiner relies on GOminer input and on CIM miner for visualization. In particular, RedundancyMiner performs Fisher's exact tests for each pair of GO terms in the datasets, calculating whether the two sets of genes annotated with these GO terms are overlapping. A symmetrical matrix of these p-values is subsequently analysed to arrive to a set of GO terms that are most independent, and therefore least redundant.

To perform the analysis, we started with the same file as for the two tools described above. First, we generated two files using a custom Python script: 1) a file containing all the genes in the array and 2) a file containing the genes that are over or underexpressed, labelled with "1" or "-1," respectively. Of note, Python is not necessary for RedundancyMiner and these files could be generated otherwise. Second, we put these files as input for the GOminer tool (http://discover.nci.nih.gov/gominer/GoCommandWebInterface.jsp). We selected the databases that contain *Homo sapiens* data and as the organism we set *H. sapiens*. The remaining parameters were the defaults in the tool. Third, we used the resulting folder as the working folder for RedundancyMiner and we ran the analysis in default mode. Finally, we visualized the resulting CIM file using cimMiner *(28)*, available at http://discover.nci.nih.gov/cimminer/home.do, in single matrix mode.

Results of our example analysis are shown in Figure 5. Even with the stringent threshold of requiring the log2 fold change greater than 5, similar trends in significant GO terms are visible as shown with the remaining two tools.

# 4 Choice of visualization

Above, we have outlined some of the currently available software tools that can visualize a set of GO terms. We have also argued that a good visualization is an effective means of discovering underlying trends in the data, while avoiding biases; an appropriate visual display is also imperative when communicating the results to others. The question of which software tool to apply should be addressed keeping these goals in mind. A related yet distinct question is which specific visualization method to choose. Here, we give a summary of the available options. Of note, the authors of this text are also the developers of REVIGO *(19)*, a versatile visualization tool, which implements several of the approaches listed below.

1. Graphs / networks. The GO graph consists of *nodes* (here, Gene Ontology terms) and *edges* (here, parent-child relationships), which connect the nodes and which may have directionality. Nodes and edges can have multiple attributes that can be visualized. For instance, the enrichment of a GO term in a user's experiment may be shown as a color of a node (Figure 2). Importantly, the spatial arrangement of the nodes on the final plot is called a *layout*, and is often created to suggest related clusters of nodes by placing similar nodes closer together. Such approaches are reviewed and demonstrated by Merico et al. *(29)*; tools like Cytoscape *(8)* support a variety of visual layouts.
   a. A special case of a layout is a tree-like display that highlights the 'levels' in the Gene Ontology and the parent-child relationships between terms (e.g., Figure 2). These levels (determining the *depth* of a node in the graph) are often used as a measure for how general the GO term is. However, this may be misleading in some instances—for example, the Molecular Function ontology is much more "shallow" than the Biological Process ontology—and we therefore recommend the use of the *information content* (IC) measure *(30)* for this purpose. This is defined as the negative logarithm of the relative frequency of the respective term annotations in some underlying database, such as the UniProt-GOA *(31)*.

2. <u>Semantic similarity space.</u> Various mathematical methods measure the semantic similarity between pairs of GO terms, such as SimRel *(32)*; see *(33)* for a review. If the first term in a pair is a direct parent, child or sibling of the second term, their semantic similarity will be very high. However, also the more distantly related terms will show some degree of similarity, as long as they reside in a common branch of the GO tree structure. Many such pairwise similarities within a group of GO terms can be processed by a projection technique, such as *principal components analysis* (PCA) or *multidimensional scaling*. The resulting plots preserve as much of the original pairwise distances as possible, while showing all supplied GO terms in a two-dimensional plane. The main visualization in REVIGO is based on this approach (Figure 3, panel A).

3. <u>Treemaps.</u> Hierarchical diagrams consisting of tiles subdivided into smaller tiles. Treemaps are good for interactive exploration, as they can be 'zoomed in' by clicking a tile and revealing finer levels of subdivisions. Here, tiles can be GO terms and the subdivisions their child terms. The tile sizes may correspond to some measure of importance of GO terms to the user, such as enrichment or p-values. REVIGO has an implementation of this visualization approach (Figure 4, panel B).

4. <u>Word clouds.</u> A display with text shown in various sizes and possibly colors. Here, the individual words or short phrases may be the names of the GO terms or some keywords associated to the GO terms. The text size/color may convey the importance to the user (enrichment), or perhaps generality of a GO term (see *information content* above). This visualization method is implemented in GOSummaries and REVIGO (Figure 4, panel C).

5. <u>Clustered Heatmaps.</u> Two-dimensional grids of values, wherein the rows and/or columns are clustered to reveal the 'block structure' in the data, clustered heatmaps are often used for showing high-dimensional data in biology, but rarely so for GO terms. In fact, this could be done to show the GO terms' similarity based on what genes are annotated to them, or on the terms' semantic similarity (which is defined by the structure of the GO graph). An example implementation can be found in RedundancyMiner (Figure 5).

In addition to the above, many of the tools specializing in GO enrichment testing (or in other analyses of large-scale biological data) often come bundled with visualizations that may include GO as an important context. Examples include the Bioconductor packages GOexpress, GOfunction and GOSim. In addition, it is often possible to customize such displays in more detail by manually passing the GO data to a dedicated visualization software, such as the *ggplot2* package *(34)* in R, or to gnuplot software. For example, a specialized software to draw treemaps can be made to display GO enrichments from a biological experiment via a script that prepares the data in a correct format *(35)*. REVIGO will draw bubble charts where the GO terms are displayed in a semantic similarity space *(19)*, and it can export a *ggplot2* script which is further customizable for e.g., font sizes, colors, and line styles; it can similarly export a graph to be further customized in Cytoscape.

## 5 Concluding remarks and outlook

In summary, we have outlined several tools that biologists can use to visualize sets of Gene Ontology terms and uncover novel and interesting trends in their experimental data. We anticipate that the future will bring even more massive biological data sets, which will have several consequences. First, the lists of interesting GO terms will grow in length, as larger sample sizes afford more statistical power to detect associations. Therefore, refinements of the existing approaches that address redundant GO terms *(19, 20)* will come in useful. Second, the visualization software will need to deal with more than a single list of enriched GO terms. While some current tools can display such results from multiple experiments side-by-side, eg. BACA *(25)*, tools will be needed that can integrate such lists and extract patterns across them. Finally, while GO is a prominent example of an ontology used by biologists, it's far from the only one *(36)*—over 100 biomedical ontologies exist that describe e.g., environments, phenotypes, and chemical entities (see chapter "Beyond the GO"). We foresee substantial developments in the tools that can summarize and visualize results of various biological experiments in the context of such emerging ontologies.

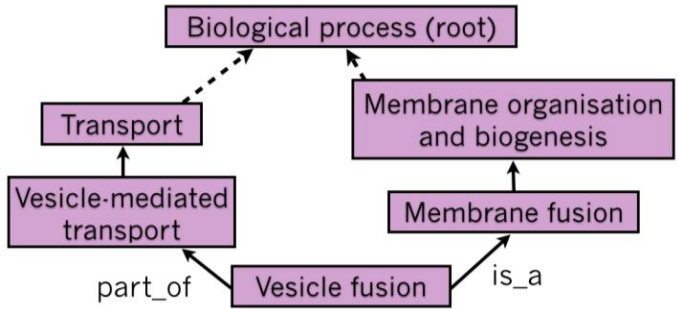

**Figure 1. A subset of the Gene Ontology Directed Acyclic Graph (DAG) for the GO term "vesicle fusion" (GO:0006906).** The GO is a DAG: terms are nodes, while the relations are edges. Two main relation types between terms are "is_a" and "part_of." More specific terms are found deeper in the graph. Thus, if a gene product is annotated with a GO term, it is by definition annotated with all the parent terms of that GO term.

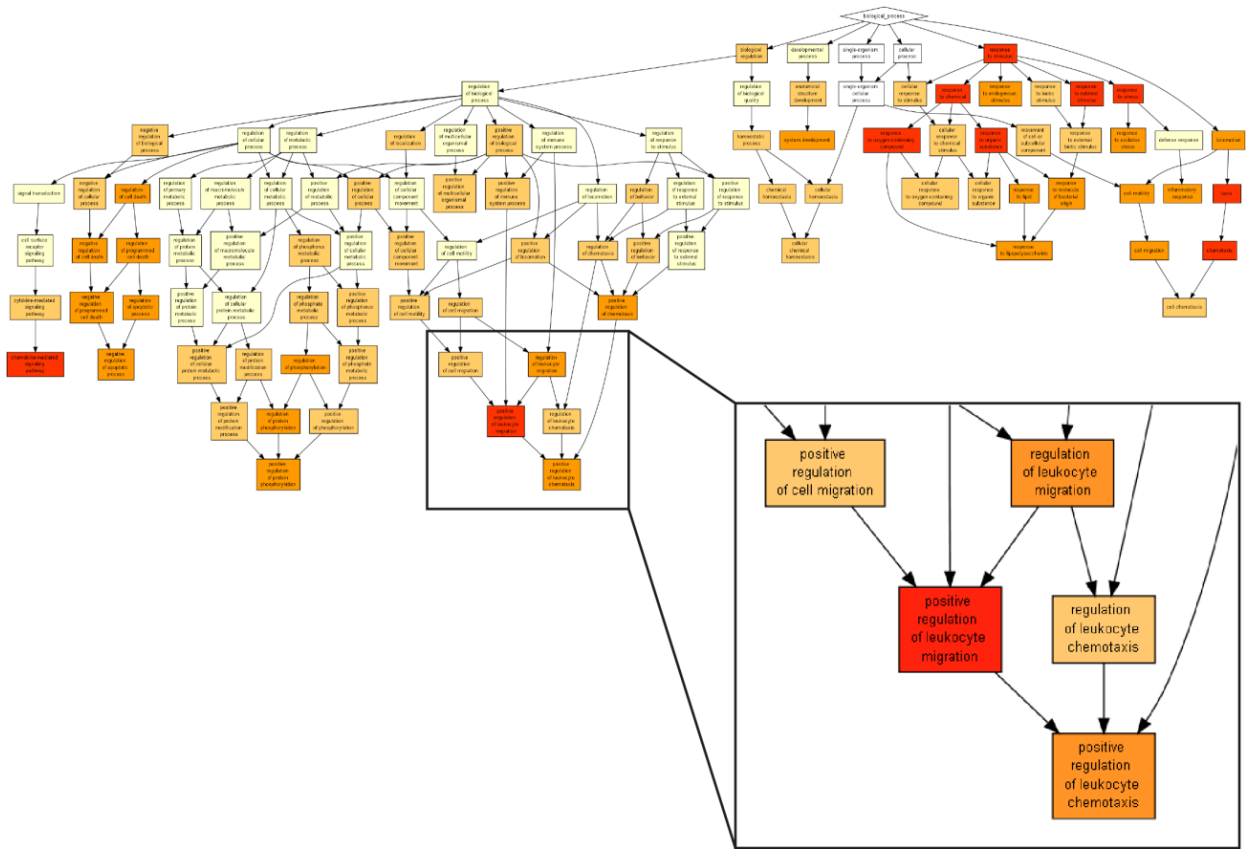

**Figure 2. A visualization of the Biological Process Gene Ontology annotations using GOrilla.** The dataset used is a microarray transcription profiling of human peripheral blood mononuclear cells after treatment with *Staphylococcus aureus* (Expression Atlas dataset ID E-GEOD-16837). The GOrilla settings were left at default values: p-value threshold of $p<10^{-3}$, organism *Homo sapiens* and running mode "single ranked list".

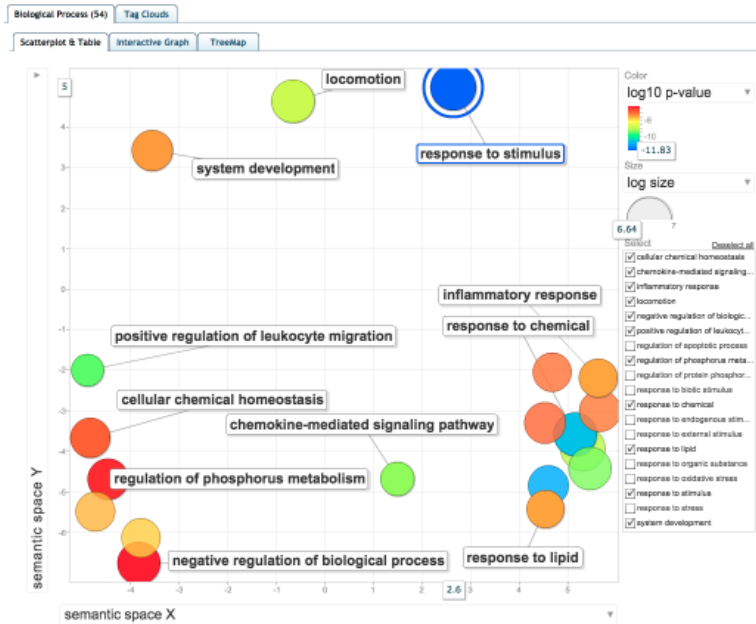

**Figure 3. Visualizations of Biological Process GO annotations using REVIGO: scatterplot and table views.** The dataset used was imported from GOrilla (see legend of Figure 2). We used the default settings of the REVIGO tool. A) The scatterplot view visualizes the GO terms in a "semantic space" where the more similar terms are positioned closer together *(19)*. The colour of the bubble reflects the p-value obtained in the GOrilla analysis, while its size reflects the generality of the GO term in the UniProt-GOA database. B) The table view shows the list of all the input GO terms: those shown in the scatterplot are written in regular font, while those labeled as redundant by REVIGO are shown in gray italics.

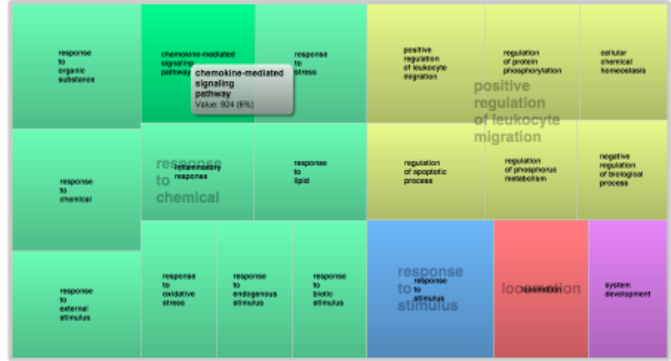
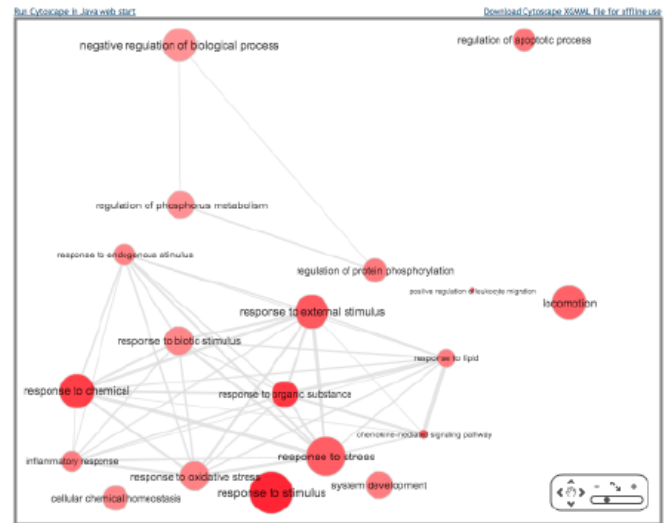
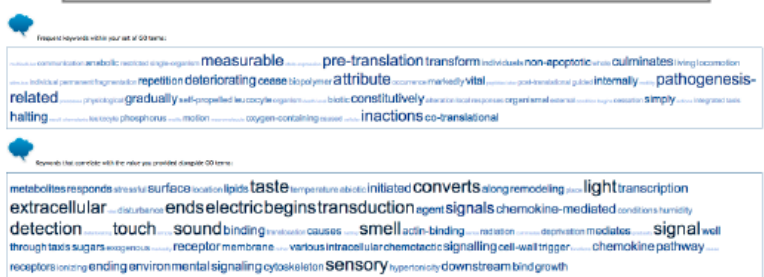

**Figure 4. Visualizations of Biological Process GO annotations using REVIGO: TreeMap (A), interactive graph (B) and word cloud views (C).** The dataset used was imported from GOrilla (see legend of Figure 2). We used the default settings of the tool.

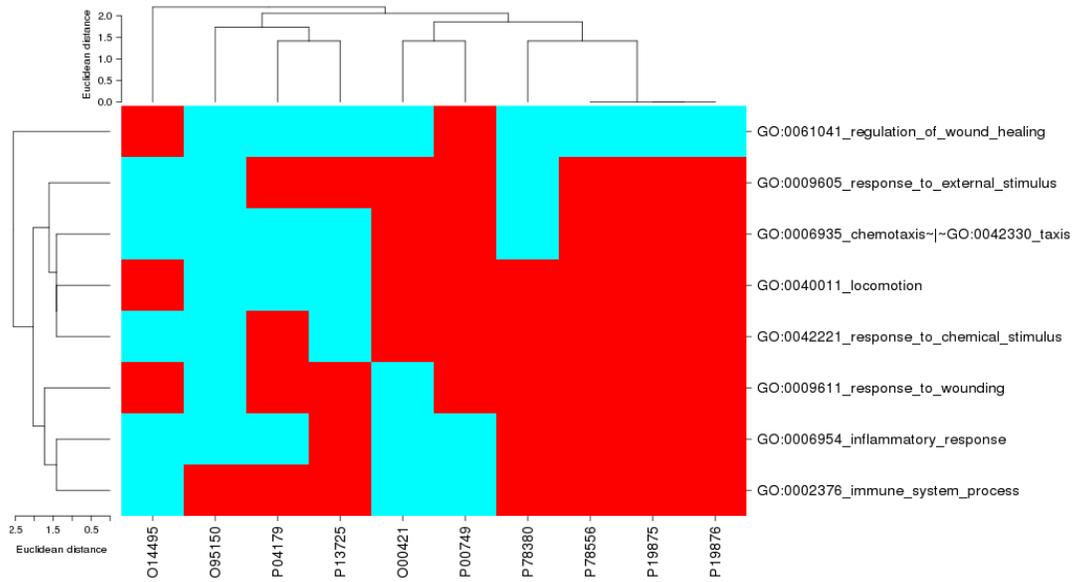

**Figure 5. A visualization of a set of Biological Process GO annotations using RedundancyMiner.** The dataset used is a microarray transcription profiling of human peripheral blood mononuclear cells after treatment with *Staphylococcus aureus*. For this visualization, we focus on genes that had $\log_2$ fold change greater than 5.